%% file: main.tex
\documentclass[letterpaper,10pt]{article}
\usepackage{style/osameet3}

\usepackage[acronym,nomain]{glossaries}
\usepackage{amsmath,amssymb}
\usepackage{textgreek}
\usepackage{xfrac}

\input{myPackages.tex}
\input{myFigures.tex}

\input{myStyles.tex}
\input{myMacros.tex}

\usepackage{subcaption}
\usepackage{wrapfig}
\usepackage{booktabs, cellspace, hhline}
\usepgfplotslibrary{fillbetween}

\usepackage[utf8]{inputenc}
\usepackage[T1]{fontenc}

\usepackage{tikz} 
\usepackage{pgfplots} 

\pgfplotsset{
    compat = 1.16, 
}

\usepackage{tikz-network} 

\usetikzlibrary{arrows.meta} 
\usetikzlibrary{calc} 

\usepackage{colortbl} 

\usetikzlibrary{backgrounds} 

\input{colors.tex}

\usetikzlibrary{arrows}
\usetikzlibrary{3d}

\usepackage{txfonts}

\input{acronyms.tex}

\begin{document}

\title{Shaped Four-Dimensional Modulation Formats for Optical Fiber Communication Systems}

\author{
Bin~Chen\textsuperscript{(1),(2)}, Gabriele Liga\textsuperscript{(2)}, Yi Lei\textsuperscript{(1),(2)}, Wei Ling\textsuperscript{(1)}, Zhengyan Huan\textsuperscript{(1)},  Xuwei Xue\textsuperscript{(3)}
and~ Alex~Alvarado\textsuperscript{(2)}}
\address{
  \textsuperscript{(1)}School of Computer and Information Engineering, Hefei University of Technology, Hefei, China\\
  \textsuperscript{(2)} Department of Electrical Engineering, Eindhoven University of Technology,  Eindhoven, the Netherlands\\ 
  \textsuperscript{(3)} Beijing University of Posts and Telecommunications, Beijing, China\\
email:{bin.chen@hfut.edu.cn}}

\begin{abstract}
We review the design of multidimensional modulations by maximizing generalized mutual information and compare 
the maximum transmission reach of recently introduced 4D formats. A model-based optimization for nonlinear-tolerant 4D modulations is also discussed.
\end{abstract}

\section{Introduction}

The design of multidimensional (MD) modulation formats has been considered as 
an effective approach to harvest performance gain in optical communications.
For an additive white Gaussian noise (AWGN) channel, higher achievable information rates are to be expected from MD shaping when increasing the constellation dimensionality \cite{ForneyJSAC1984}.
On the other hand, nonlinear
effects in the optical channel could be mitigated by MD  geometrical shaping \cite{Dar14_ISIT}.
This insight motivates the search for a linear noise and /or nonlinear interference (NLI)-tolerant modulation formats in a higher dimensional space.

Conventional MD formats are not \textit{true} MD formats in the sense that they are only optimized in each dimension independently. This is the case of  polarization-multiplexed 2D (PM-2D) formats. 
MD modulation formats with dependency between dimensions can be obtained by set-partitioning the regular QAM (SP-QAM) or via MD  geometrical shaping. 
MD coded modulation encodes binary bits and then maps them onto consecutive 2D symbols or onto one MD symbol. For example in 4D space, polarization switched-QPSK \cite{Karlsson:09} maps 3 bits onto two consecutive QPSK symbols while 4D2A8PSK \cite{Kojima2017JLT} maps 5-7 bits onto two 8PSK symbols in both X and Y polarizations.
More recently, modulation formats in 4D, 8D and 12D have been proposed by adding constraints in the  optimization to enable larger gains in nonlinearity tolerance and to further extend the transmission reach \cite{BinChenJLT2019,BinChenPTL2019,ReneECOC2020}.

In this paper, we focus on designing 4D modulation formats for soft-decision forward error correction (SD-FEC)  with 20\%-25\% overhead by maximizing the generalized mutual information (GMI) and ,thus, increase transmission reach.  
Simulation comparisons for a set of 4D-optimized modulation formats, which outperform previously known 4D formats, are presented.
Finally, to highlight future directions for the design of nonlinear-tolerant modulation in optical fiber systems, an optimization of dual-polarization (DP) modulation based on 4D NLI model  \cite{GabrieleEntropy2020} 
is performed.

\section{GMI Computation and Design Methodology for Multidimensional Modulation}

\begin{wrapfigure}{r}{0.55\textwidth}
\vspace{-2.8em}
\centering

\includegraphics[width=0.55\textwidth]{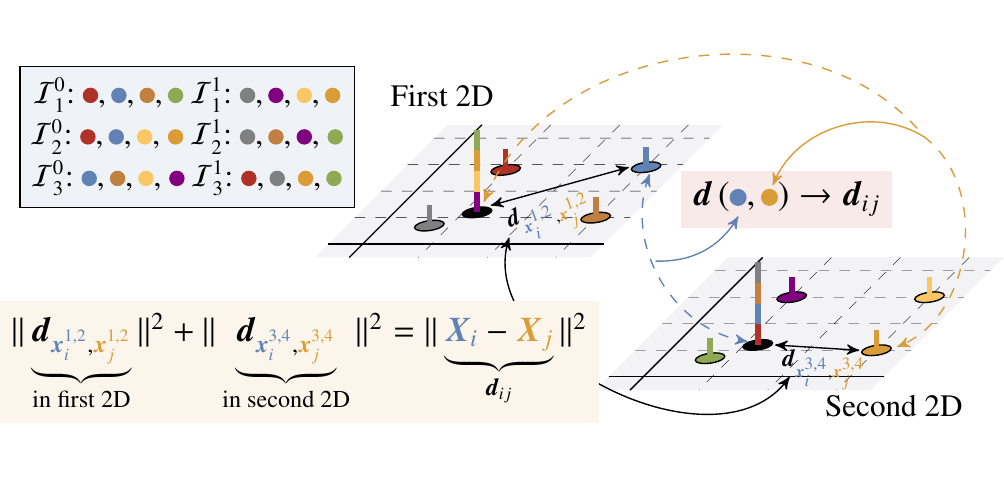}
\vspace{-2.5em}
\caption{Example of  $\bdij$ calculation and the constellation set $\mcIkb$  for  4D-OS128 \cite{BinChenJLT2021} in first orthant of 4D space (2$\times$2D).} 
\label{fig:3D}
\vspace{-0.5em}
\end{wrapfigure}

Due to its simplicity and flexibility, bit-interleaved coded modulation with SD-FEC is usually considered an attractive option for optical fiber communication systems \cite{SmithJLT2012}, and hence, the use of information-theoretical performance metric GMI is preferred for coded modulation design \cite{AlvaradoJLT2018}.

For a  discrete uniformly-distributed $N$-dimensional modulation  with spectral efficiency (SE) $m=\log_2M$ bit/4D, the GMI under Gaussian noise assumption can be estimated via Gauss-Hermite quadrature as \cite [Eq. (45)]{AlvaradoJLT2018},
\vspace{-0em}
\begin{align}\label{eq:GMI}
\text{GMI} \approx m-  \frac{1}{M\pi^{N/2}}\sum_{k=1}^{m}\sum_{b\in\set{0,1}}\sum_{i\in\mcIkb}\sum_{l_1=1}^{\GHs}\alpha_{l_1}\sum_{l_2=1}^{\GHs}\alpha_{l_2}\cdots\sum_{l_N=1}^{\GHs}\alpha_{l_N} \cd \log_{2}\frac{\sum_{p=1}^{M}\exp\left(-\frac{||\bdip||^{2}+2\sigma_{z}\Re\set{(\xi_{l_1}+\jmath\xi_{l_2})\bdip}}{\sigma_{z}^{2}}\right)}{\sum_{j\in\mcIkb}\exp\left(-\frac{||\bdij||^{2}+2\sigma_{z}\Re\set{(\xi_{l_1}+\jmath\xi_{l_2})\bdij}}{\sigma_{z}^{2}}\right)},
\end{align}
where the quadrature nodes $\xi_l$   and the weights $\alpha_l$ can be easily found (numerically) for different values of $\GHs$. In this paper, we use  the  quadrature nodes and weights for $J=10$ in  \cite [Table III]{AlvaradoJLT2018}.
$\bdij \triangleq\bX_i-\bX_j$ denote  the difference between two MD symbols, $\bX_i=[\bx_i^{1,2},\bx_i^{3,4},\cdots,\bx_i^{N-1,N}]$  denotes a MD symbol consisting of $N/2$ complex symbols, $\sigma_{z}^{2}$ is the noise variance per complex dimension and  $\mcIkb\subset\set{1,2,\ld,M}$ with $|\mcIkb|=M/2$ is the set of indices of constellation points whose binary label is $b$ at bit position $k$.  Fig.  \ref{fig:3D} shows an example of computing $\bdij$ of two 4D symbols as $\boldsymbol{d}\left(\tikz\draw[blue,fill=blue] (0,0) circle (.5ex);, \tikz\draw[orange,fill=orange] (0,0) circle (.5ex); \right)$ for  4D  format 4D-OS128.
In order to clearly show the  dependency of 4D symbols, 
we use a similar color
coding  as in \cite{BinChenJLT2021}: valid 4D symbols are the 
2D projected symbols in the first/second 2D with the same color.

As shown in Eq. \eqref{eq:GMI}, GMI computation requires a joint consideration of the 4D coordinates and its binary labeling. A GMI-based optimization  can find a constellation $\mathcal{X}^*$ and a labeling  $\mathcal{L}^*$  for a given channel conditional PDF $p_{\bY|\bX}$ with a constraint on transmitted power $\sigma^2_x$, i.e.,
$\{\mathcal{X}^*,\mathcal{L}^*\}  = \argmax_{\mathcal{X},\mathcal{L}: E[\|\bX\|^2]\leq \sigma_x^2} G(\mathcal{X},\mathcal{L},p_{\bY|\bX})$, 
where $G$ as an  expression of  GMI  emphasizes the dependency of the GMI on the constellation, binary labeling, and channel law.

It is known that GMI-based optimization of large constellations and/or constellations with high dimensionality is computationally demanding.
Therefore, an unconstrained optimization with at least hundreds of GMI evaluations is very challenging. Potentially irregular formats obtained from the optimization also impose strict requirements on the generation and detection of the signals, due to the need of high-resolution digital-to-analog/analog-to-digital converters.
To solve the multi-parameter optimization challenges of MD geometric shaping and also to  achieve a good performance-complexity tradeoff, constraints of constant modulus \cite{Kojima2017JLT,BinChenJLT2019} and orthant-symmetry (OS) \cite{BinChenJLT2021}   
have been proposed to design $N$-dimensional  formats. These solutions have shown a small performance loss with respect to the unconstrained optimizations in AWGN channel and achieve an even better performance in the optical fiber channel.

\begin{wraptable}{r}{0.32\textwidth}
\vspace{-2.5em}
\caption{Simulation parameters.}\label{tab:para}
\vspace{-0.5em}
\begin{footnotesize}
\begin{tabular}{c|c}
\hline\hline
\multicolumn{2}{c}{\textbf{TX Parameters}} \\
\hline
Symbol rate & 45 Gbaud \\
No. of WDM channels & 11 \\
Channel spacing & 50 GHz \\
Root-raised-cosine roll-off & 10\% \\
\hline
\multicolumn{2}{c}{\textbf{Fiber and Link Parameters}} \\
\hline
Attenuation coeff. ($\alpha$) & 0.21 dB/km \\
Disp. parameter ($D$) & 16.9 ps/nm/km \\
Nonlinear coeff. ($\gamma$) & 1.31 dB/km \\
Span length & 80 km \\
EDFA noise figure & 5 dB \\
\hline\hline
\end{tabular}
\end{footnotesize}
\vspace{-1em}
\end{wraptable}

\section{Simulation Results of 4D Geometric Shaping for Multi-Span Systems}

 To target a practical SD-FEC  with 20\%-25\% overhead, the optimizations were performed for AWGN channel at an SNR in which $\text{GMI}\approx0.85m$ for six different SEs with $m\in\{5,6,7,8,9,10\}$. 
{In order to make the modulation more structured and reduce the optimization complexity, constraint of OS is used for $m=7, 8, 9, 10$.}
Split-step Fourier method of the nonlinear Manakov equation with a step size of 100 m was performed to compare the
modulation formats and predict system performance. The simulation parameters are given in Table \ref{tab:para} for the 
optical fibre link under consideration.

\begin{figure}[!htbp]
		\vspace{-0.2cm}
\centering
\begin{subfigure}{0.63\textwidth}
\includegraphics[width=1.05\textwidth]{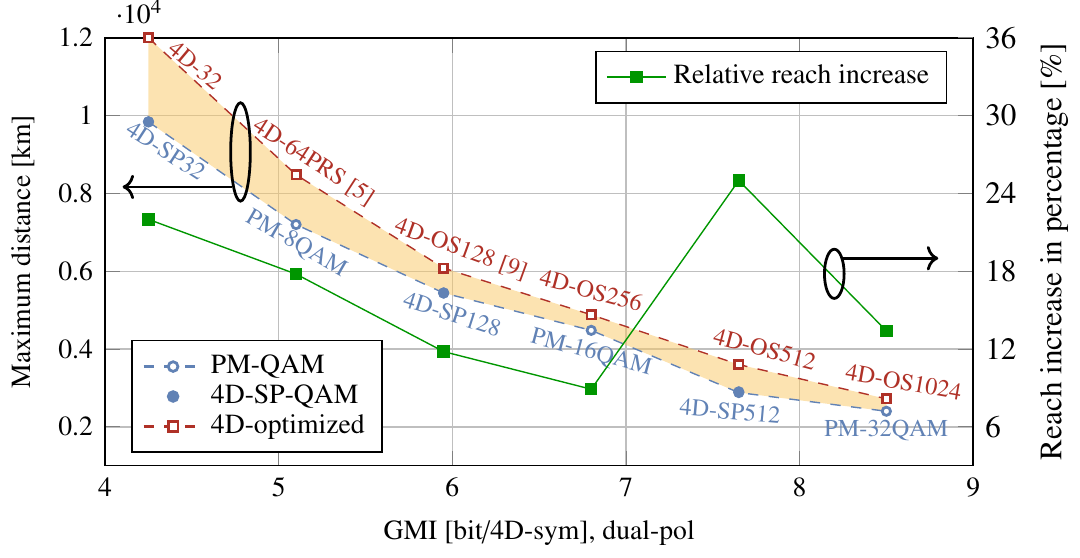}
\end{subfigure}
\begin{subfigure}{.36\textwidth}

\vspace{-0.7em}
{\hspace{5em}  \footnotesize (a)   \hspace{8em}  \footnotesize (d) }
\vspace{-0.3em}

{\hspace{4.5em}  \footnotesize 4D-32
\hspace{6em}  4D-OS256}

\hspace{2em}\includegraphics[width=0.215\textwidth]{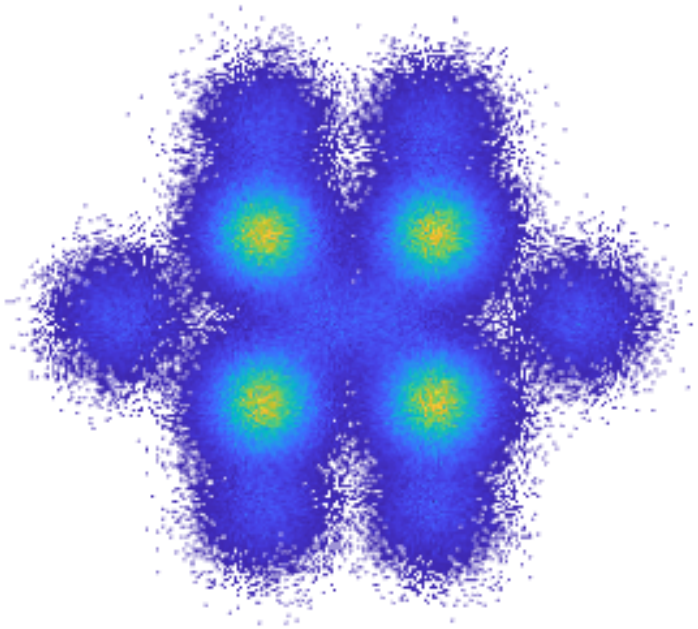}\hspace{-0.3em}
\includegraphics[width=0.215\textwidth]{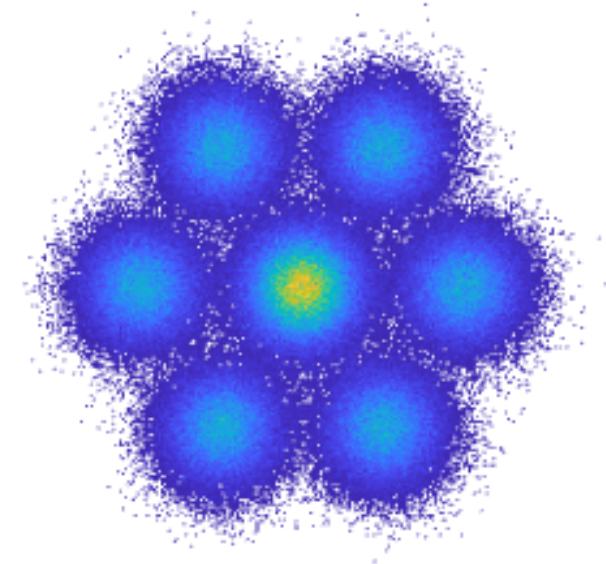} \hspace{0.2em}
\includegraphics[width=0.2\textwidth]{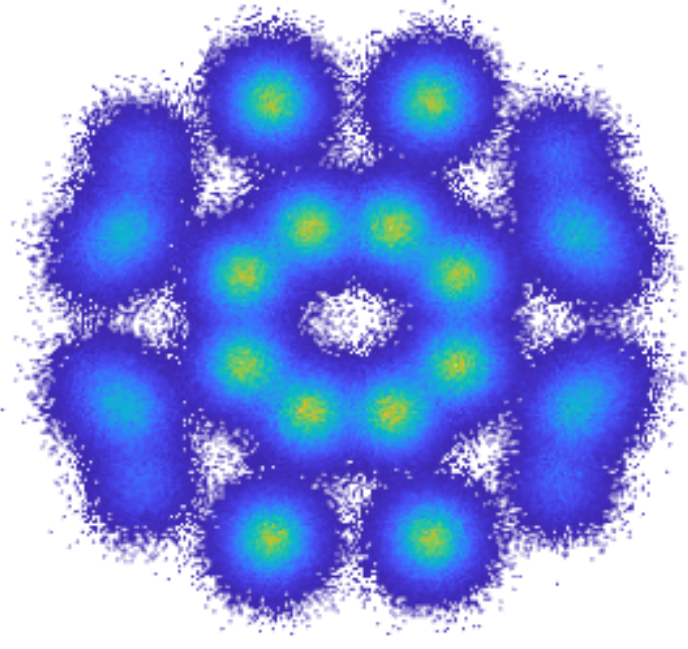}\hspace{0.1em}
\includegraphics[width=0.2\textwidth]{figure/256_apsk_os_constellation_noise_15dB_X.pdf}

\vspace{-0.6em}

{\hspace{5em}  \footnotesize (b)   \hspace{8em}  \footnotesize (e) }
\vspace{-0.3em}

{\hspace{4em}\footnotesize 4D-64PRS
\hspace{5em}  4D-OS512}

\hspace{2em}\includegraphics[width=0.2\textwidth]{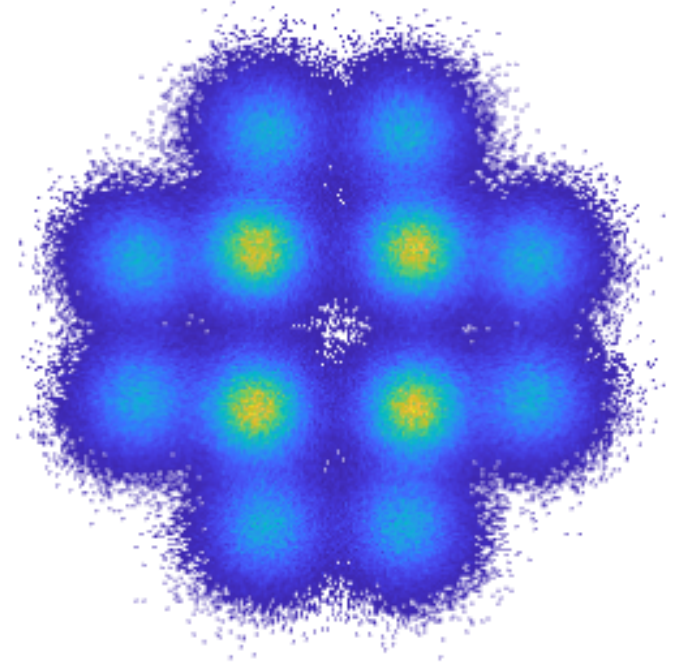}\hspace{0.1em}
\includegraphics[width=0.2\textwidth]{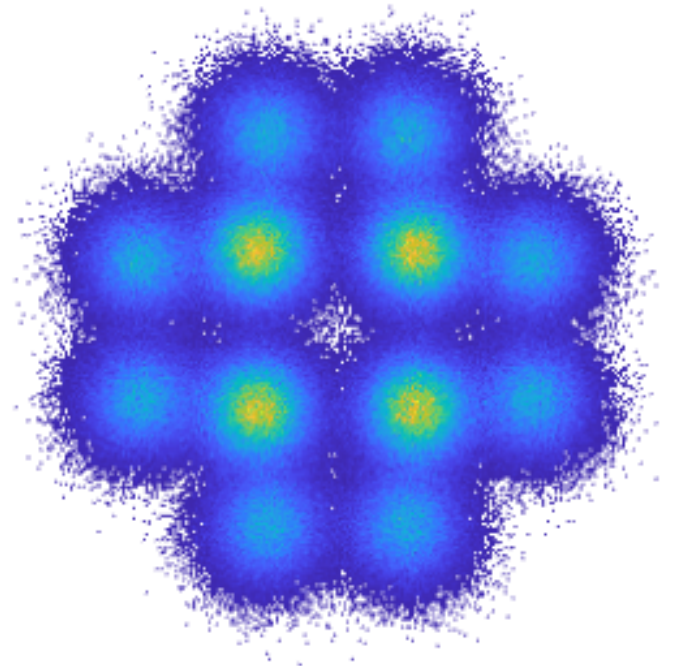} \hspace{0.2em}
\includegraphics[width=0.2\textwidth]{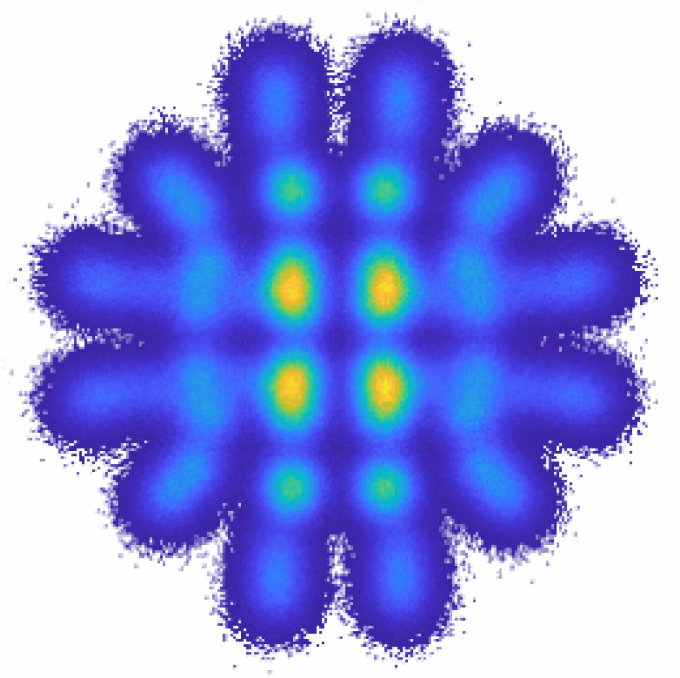}\hspace{0.1em}
\includegraphics[width=0.2\textwidth]{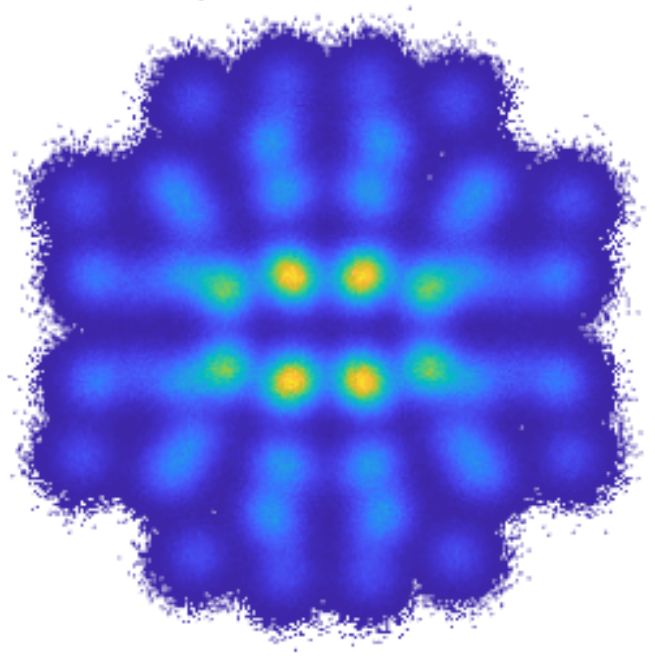}

\vspace{-0.6em}

{\hspace{5em}  \footnotesize (c)   \hspace{8em}  \footnotesize (f) }
\vspace{-0.3em}

{\hspace{4em}\footnotesize 4D-OS128
\hspace{5em}  4D-OS1024}

\hspace{2em}\includegraphics[width=0.2\textwidth]{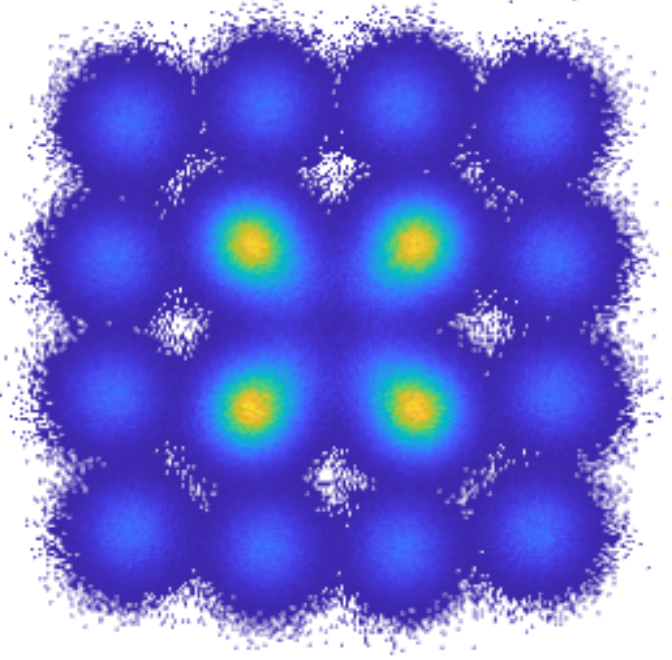}\hspace{0.1em}
\includegraphics[width=0.2\textwidth]{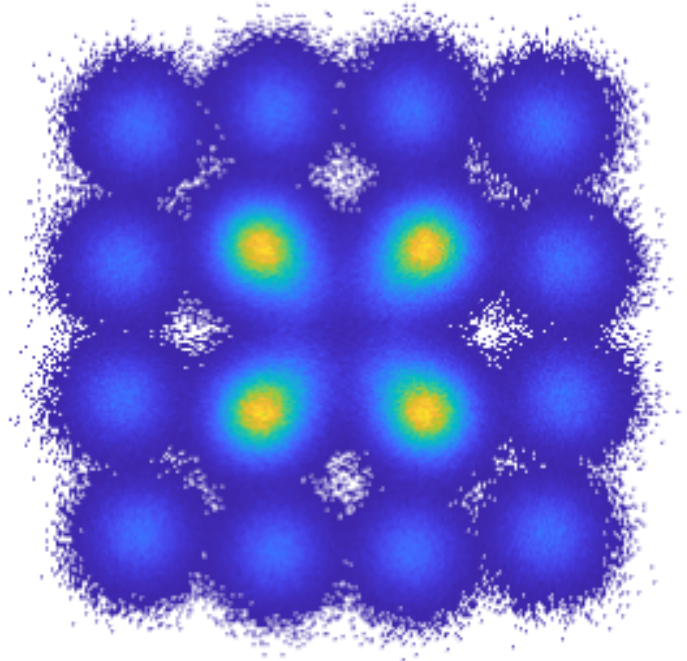} \hspace{0.2em}
\includegraphics[width=0.2\textwidth]{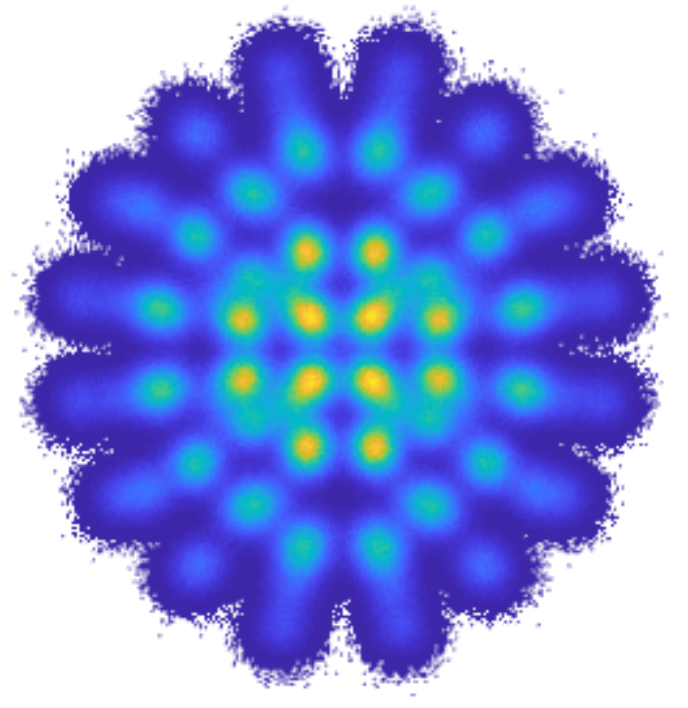}\hspace{0.1em}
\includegraphics[width=0.21\textwidth]{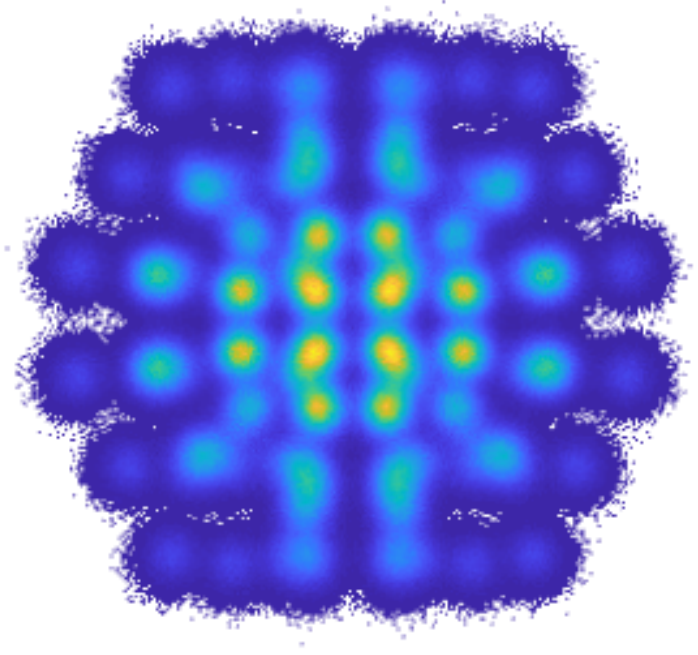}
\end{subfigure}
	\vspace{-0.2cm}
\caption{The maximum reach of various modulation formats for multi-span optical fiber transmission. 
The $2\times$ 2D projection of the modulations at normalized GMI of 0.95 are depicted as  (a) - (f).   
} 
\label{fig:DvsSE}
	\vspace{-0.5cm}
\end{figure}

In Fig. \ref{fig:DvsSE}, the maximum transmission distance and the relative reach increase in percentage at $\text{GMI}=0.85m$  of  twelve modulation formats are evaluated. 
We observe that 4D-optimized  formats achieve approximately
320-2160~km (9\%-25\%) reach increase w.r.t  PM-QAM/4D-SP-QAM at the same information rates, which are highlighted by  the orange shaded region.
We note from Fig. \ref{fig:DvsSE} that 
larger reach increase in percentage  can be achieved w.r.t the QAM  modulations  without gray labeling.  
Especially comparing to  4D-SP32 and 4D-SP512,  the gains of  4D-optimized formats are more than 20\%, which is mainly due to the superior performance  of labeling.

\section{4D  NLI Model-aided 4D Geometric Shaping for Single-Span Transmission}

As noted in the previous section, {most of the modulation formats in Fig. \ref{fig:DvsSE} are  designed for AWGN channel, only 4D-64PRS uses heuristic idea of constant-modulus constraint to improve the nonlinearity tolerance.}  
For nonlinear fiber channel, NLI power models with considering modulation-dependent interference could provide a quick computation of the NLI power as a function of the input constellation, e.g., the enhanced Gaussian noise model \cite{Carena:14} for  PM-2D format and 4D NLI model \cite{GabrieleEntropy2020} for a general DP-4D format.
Accordingly, to design a nonlinear-tolerant 4D modulation, the optimization problem for a given optical fiber channel parameters $\mathcal{P}$ can be reformulated as,

\vspace{-0.5cm}
\begin{align}\label{eq:opt_4Dmodel}
	\{\mathcal{X}^*,\mathcal{L}^*\} & = \argmax_{\mathcal{X},\mathcal{L}
	} G\left(\mathcal{X},\mathcal{L},\text{SNR}_{\text{opt}}(\mathcal{X},\mathcal{P})\right),
\end{align}
where $\text{SNR}_{\text{opt}}(\mathcal{X},\mathcal{P})$ denotes the optimum effective SNR at a given distance and  depends on the modulation format.  

\begin{figure}[!htbp]
	\vspace{-0.2cm}
\centering
\includegraphics[width=0.51\textwidth]{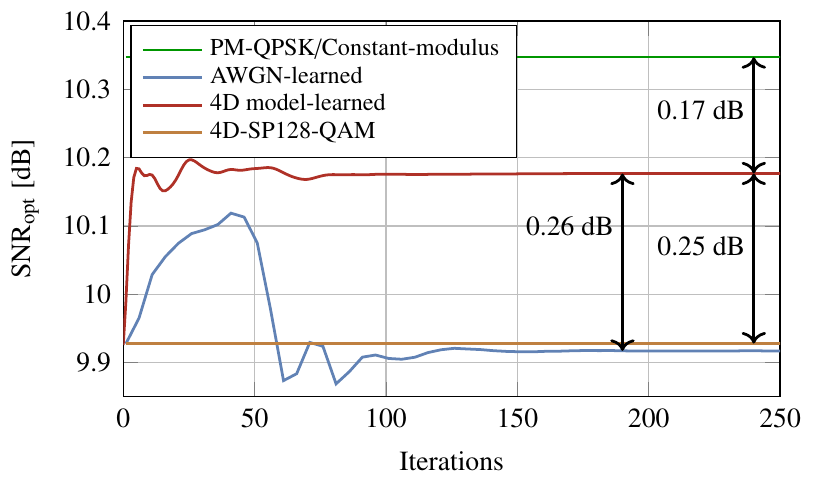}\includegraphics[width=0.51\textwidth]{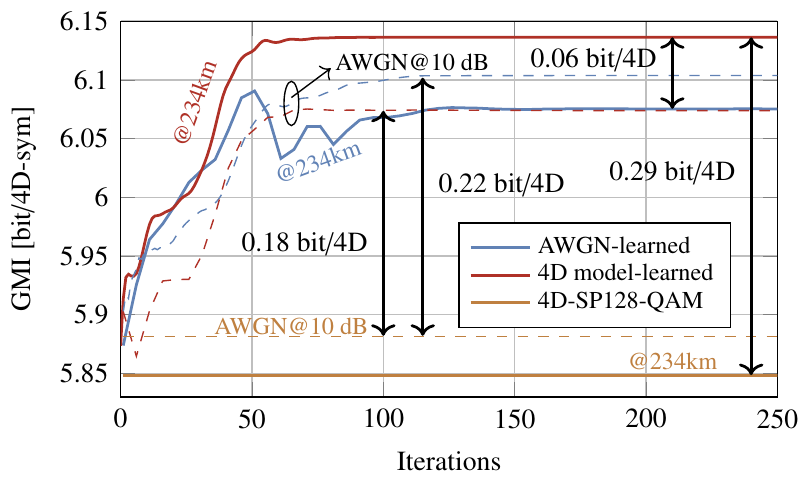}
	\vspace{-0.4cm}
\caption{The  $\text{SNR}_{\text{opt}}$ at 234~km (left) and GMI (right) of geometrically shaped 4D modulation format optimized based on AWGN channel and 4D NLI model. In this example, $M=128$.  
} 
\label{fig:4Dmodel_opt}
	\vspace{-0.6cm}
\end{figure}

In Fig.  \ref{fig:4Dmodel_opt}, the 4D modulation  formats with a SE of 7~bit/4D  are optimized with OS constraint via end-to-end learning following \cite{Kadir2019endtoend} by maximizing GMI.
The  simulations  are  implemented  by  solving two  optimization  problems: one is for AWGN channel with SNR=10~dB (AWGN-learned) and the other is for the 4D-model \cite{GabrieleEntropy2020} with a single-channel, 234km single-span transmission system (4D model-learned). 
PM-QPSK as a format with a good nonlinearity tolerance and 4D-128SP-QAM \cite{ErikssonOE13} with 7~bit/4D  are shown as references.

Fig.~\ref{fig:4Dmodel_opt} (left) shows that the 4D model-learned modulation can tolerate higher nonlinearity that achieves up to 0.25~dB gain with respect to 4D-SP128-QAM in  terms  of  $\text{SNR}_{\text{opt}}$  at 234~km. 
Fig.  \ref{fig:4Dmodel_opt} (right) shows that in an AWGN channel with an SNR of 10~dB,  AWGN-learned modulation format provide the gain around 0.22~bit/4D (in term of GMI)  with respect to 4D-SP128-QAM, while the gain is around 0.18~bit/4D for 4D-learned modulation format. However, the gain of the 4D model-learned modulation in an optical channel with fiber length of 234~km is increased to 0.29~bit/4D, which is higher than that of the AWGN-learned format. This benefits from the improvement of $\text{SNR}_{\text{opt}}$ shown in Fig.~\ref{fig:4Dmodel_opt} (left).  
It well indicates that 4D model-learned modulation  leads to a good trade-off between linear and nonlinear shaping gain  {by  increasing the linear shaping gain and  maintaining a fair level of nonlinearity tolerance.}

\section{Conclusions}

We numerically assessed a series of multidimensional 
modulation formats for multi-span transmission systems. We showed that the 4D-optimized modulation formats can be a solution for multi-rate applications between 5 and 10~bit/dual-pol.
In addition, up to 0.25~dB NLI gains in terms of $\text{SNR}_{\text{opt}}$  are demonstrated for 4D model-based modulation optimization over a  regular 4D format for a single-span transmission system. 
 The results in this work confirm that the multidimensional modulations could be a good alternative for high capacity  transmission systems and offer  substantial potential gains in the nonlinear optical fiber channel.
 
\begin{footnotesize}
\textbf{Acknowledgements:} {The work of B. Chen and  Y. Lei are supported by the National Natural Science Foundation of China (NSFC)
under Grant 62171175 and 62001151. 
The work of G.~Liga and A. Alvarado are funded by the EuroTechPostdoc programme and the European Research Council under the European Union’s Horizon 2020 research and innovation programme (grant agreement No. 754462 and  757791).
\par}
\end{footnotesize}


\small
\bibliographystyle{style/osajnl}
\bibliography{references}

\end{document}

%% file: myPackages.tex
\usepackage[utf8]{inputenc}

\usepackage{amsmath,amssymb}

\usepackage{xcolor}
\usepackage{graphicx}
\usepackage{tikz}
\usepackage{pgfplots}
\usepackage{float}
\usepackage{balance}
\usepackage{pgfplotstable}
\usepackage{xspace}
\usepackage[normalem]{ulem}

\usepackage{dsfont,empheq}
\usepackage{multicol,multirow}

\usepackage{cite}

%% file: myFigures.tex
\usetikzlibrary{plotmarks,matrix,chains,scopes,fit,calc,shapes,positioning,decorations,intersections,arrows,backgrounds,shadows}
\usetikzlibrary{fit}
\usetikzlibrary{shapes.multipart}
\usetikzlibrary{positioning}
\usetikzlibrary{shapes}
\usetikzlibrary{arrows.meta}

\newlength\FigureHeight
\newlength\FigureWidth


%% file: myStyles.tex
\definecolor{myDarkGreen}{rgb}{0.00000,0.58824,0.00000}%
\definecolor{uniform}{rgb}{0.00000,0.58824,0.00000}%
\definecolor{AWGNreference}{rgb}{0.00000,0.58824,0.00000}%

%% file: myMacros.tex
\DeclareMathOperator*{\argmax}{argmax}

\newcounter{lemma}

\newtheorem{exampleplain}{Example}

\newcommand{\cd}{\cdot}
\newcommand{\mcIkb}{\mathcal{I}_{k}^{b}}
\newcommand{\bdij}[0]{\boldsymbol{d}_{ij}}

\newcommand{\bdip}[0]{\boldsymbol{d}_{ip}}

\newcommand{\set}[1]{\{#1\}}

\newcommand{\ld}{\ldots}

\newcommand{\bX}{\boldsymbol{X}}\newcommand{\bx}{\boldsymbol{x}}

\newcommand{\bY}{\boldsymbol{Y}}

\newcommand{\GHs}{J}



%% file: colors.tex
\definecolor{blue}{rgb}{0.38, 0.51, 0.71} 
\definecolor{darkblue}{RGB}{17, 42, 60} 
\definecolor{red}{RGB}{175, 49, 39} 

\definecolor{orange}{RGB}{217, 156, 55} 
\definecolor{green}{RGB}{144, 169, 84} 
\definecolor{palegreen}{RGB}{197, 184, 104} 

\definecolor{yellow}{RGB}{250, 199, 100} 
\definecolor{brokenwhite}{RGB}{218, 192, 166} 
\definecolor{brokengrey}{rgb}{0.77, 0.76, 0.82} 

%% file: acronyms.tex
\newacronym{KK}{KK}{Kramers-Kronig}
\newacronym{CSPR}{CSPR}{carrier-to-signal power ratio}
\newacronym{KKRX}{KKRx}{Kramers-Kronig Receiver}
\newacronym{SSBI}{SSBI}{signal-signal beat interference}

\newacronym{DSP}{DSP}{digital signal processing}
\newacronym{MIMO}{MIMO}{multiple-input multiple-output}
\newacronym{TDE}{TDE}{time domain equalizer}
\newacronym{FDE}{FDE}{frequency domain equalizer}
\newacronym{LMS}{LMS}{least means square}
\newacronym{DDLMS}{DD-LMS}{decision directed least means square}
\newacronym{FFE}{FFE}{feed-forward equalizer}
\newacronym{FBE}{FBE}{feedback equalizer}

\newacronym{SMF}{SMF}{single-mode fiber}
\newacronym[plural=SSMFs]{SSMF}{SSMF}{standard single-mode fiber}
\newacronym[plural=FMFs]{FMF}{FMF}{few-mode fiber}
\newacronym{FMF12}{FMF12}{12 mode FMF}
\newacronym{MMF}{MMF}{multi-mode fiber}
\newacronym{SI}{SI}{step index}
\newacronym{GI}{GI}{graded index}
\newacronym{DCF}{DCF}{dispersion compensated fiber}

\newacronym{SDM}{SDM}{space division multiplexing}
\newacronym{MDM}{MDM}{mode division multiplexed}
\newacronym{WDM}{WDM}{wavelength division multiplexing}

\newacronym{LP}{LP}{linear polarized}
\newacronym[plural=MMUXs,firstplural=mode multiplexers]{MMUX}{MMUX}{mode multiplexer}
\newacronym{PL}{PL}{photonic lantern}
\newacronym{3DWG}{3DWG}{3D-waveguide}
\newacronym{MDL}{MDL}{mode dependent loss}
\newacronym{DGD}{DGD}{differential group delay}
\newacronym{DMGD}{DMGD}{differential mode group delay}
\newacronym{QSM}{QSM}{quasi-single-mode}
\newacronym{GIMMF}{GI-MMF}{graded-index multi-mode fiber}

\newacronym{SSB}{SSB}{single side band}
\newacronym{QPSK}{QPSK}{quadrature phase shift keying}
\newacronym{QAM}{QAM}{quadrature amplitude modulation}
\newacronym{RRC}{RRC}{root-raised-cosine}

\newacronym{ECL}{ECL}{external cavity laser}
\newacronym{CW}{CW}{continuous wave}
\newacronym[plural=DFBs]{DFB}{DFB}{distributed feedback laser}

\newacronym[plural=DACs]{DAC}{DAC}{digital to analog converter}
\newacronym{ADC}{ADC}{analog to digital converter}
\newacronym{PRBS}{PRBS}{pseudo-random bit sequence}
\newacronym{LO}{LO}{local oscillator}
\newacronym{EDFA}{EDFA}{erbium doped fiber amplifier}
\newacronym{MZM}{MZM}{Mach-Zehnder modulator}
\newacronym{DP-MZM}{DP-MZM}{dual-polarization Mach-Zehnder modulator}
\newacronym{ChUT}{ChUT}{channel under test}
\newacronym{WSS}{WSS}{wavelength selective switch}
\newacronym[plural=VOAs]{VOA}{VOA}{variable optical attenuator}
\newacronym[plural=PDCRXs]{PDCRX}{PDCRX}{polarization diverse coherent receiver}
\newacronym{DSO}{DSO}{digital storage oscilloscope}
\newacronym{ASE}{ASE}{amplified spontaneous emission}
\newacronym{PBS}{PBS}{polarization beam splitter}
\newacronym{PD}{PD}{photodiode}

\newacronym{OSNR}{OSNR}{optical signal to noise ratio}
\newacronym{BER}{BER}{bit error rate}
\newacronym{IL}{IL}{insertion loss}

\newacronym{SDFEC}{SD-FEC}{soft decision forward error correction}
\newacronym{HDFEC}{HD-FEC}{soft decision forward error correction}
\newacronym{FEC}{FEC}{forward error correction}

\newacronym{AIR}{AIR}{achievable information rate}
\newacronym{AR}{AR}{achievable rates}
\newacronym{MI}{MI}{mutual information}

\newacronym{OVNA}{OVNA}{optical vector network analyzer}
\newacronym{NIR}{NIR}{near infrared}
\newacronym{CD}{CD}{chromatic dispersion}
\newacronym{OTDR}{OTDR}{optical time domain reflectometry}
\newacronym{OFDR}{OFDR}{optical frequency domain reflectometry}
\newacronym{GPU}{GPU}{graphics processing unit}
\newacronym{SVD}{SVD}{singular value decomposition}
\newacronym{WGN}{WGN}{white Gaussian noise}
\newacronym{AWGN}{AWGN}{additive white Gaussian noise}
\newacronym{PDL}{PDL}{polarization dependent loss}
\newacronym{SPS}{sps}{samples-per-symbol}

%% file: main.bbl
\begin{thebibliography}{10}
\newcommand{\enquote}[1]{``#1''}

\bibitem{ForneyJSAC1984}
G.~Forney \emph{et~al.}, \enquote{Efficient modulation for band-limited
  channels,} {{IEEE Journal on Selected Areas in Communications}} \textbf{2},
  632--647 (1984).

\bibitem{Dar14_ISIT}
R.~Dar \emph{et~al.}, \enquote{On shaping gain in the nonlinear fiber-optic
  channel,} in \emph{{IEEE} Int. Symp. on Inf. Theory,}  (2014), pp.
  2794--2798.

\bibitem{Karlsson:09}
M.~Karlsson and E.~Agrell, \enquote{Which is the most power-efficient
  modulation format in optical links?} {{Opt. Express}} \textbf{17} (2009).

\bibitem{Kojima2017JLT}
K.~Kojima \emph{et~al.}, \enquote{Nonlinearity-tolerant four-dimensional
  {2A8PSK} family for 5-7 bits/symbol spectral efficiency,} {{JLT}} \textbf{35}
  (2017).

\bibitem{BinChenJLT2019}
B.~{Chen} \emph{et~al.}, \enquote{Polarization-ring-switching for
  nonlinearity-tolerant geometrically-shaped four-dimensional formats
  maximizing generalized mutual information,} {{JLT}} \textbf{37} (2019).

\bibitem{BinChenPTL2019}
B.~{Chen} \emph{et~al.}, \enquote{Eight-dimensional polarization-ring-switching
  modulation formats,} {{PTL}} \textbf{31} (2019).

\bibitem{ReneECOC2020}
R.-J. Essiambre \emph{et~al.}, \enquote{Increased reach of long-haul
  transmission using a constant-power {4D} format designed using neural
  networks,} in \emph{ECOC,}  (2020).

\bibitem{GabrieleEntropy2020}
G.~Liga \emph{et~al.}, \enquote{Extending fibre nonlinear interference power
  modelling to account for general dual-polarisation {4D} modulation formats,}
  {{Entropy}} \textbf{22} (2020).

\bibitem{BinChenJLT2021}
B.~Chen \emph{et~al.}, \enquote{Analysis and experimental demonstration of
  orthant-symmetric four-dimensional 7 bit/{4D}-sym modulation for optical
  fiber communication,} {{JLT}} \textbf{39} (2021).

\bibitem{SmithJLT2012}
B.~P. {Smith} \emph{et~al.}, \enquote{A pragmatic coded modulation scheme for
  high-spectral-efficiency fiber-optic communications,} {{JLT}} \textbf{30}
  (2012).

\bibitem{AlvaradoJLT2018}
A.~Alvarado \emph{et~al.}, \enquote{Achievable information rates for fiber
  optics: Applications and computations,} {{JLT}} \textbf{36} (2018).

\bibitem{Carena:14}
A.~Carena \emph{et~al.}, \enquote{{EGN} model of non-linear fiber propagation,}
  {{Opt. Express}} \textbf{22} (2014).

\bibitem{Kadir2019endtoend}
K.~Gümüs \emph{et~al.}, \enquote{End-to-end learning of geometrical shaping
  maximizing generalized mutual information,} in \emph{OFC,}  (2020).

\bibitem{ErikssonOE13}
T.~A. Eriksson \emph{et~al.}, \enquote{Comparison of {128-SP-QAM} and
  {PM-16QAM} in long-haul {WDM} transmission,} {{Opt. Express}} \textbf{21}
  (2013).

\end{thebibliography}
